\documentclass[
aps,
prl,
showpacs,
tightenlines,
nofootinbib,
floatfix]
{revtex4}
\usepackage{graphicx,
longtable
}

\begin{document}
\title{	Hints of $\theta_{13}>0$ from global neutrino data analysis}
\author{G.L.~Fogli$^{1,2}$, E.~Lisi$^2$, A.~Marrone$^{1,2}$, A.~Palazzo$^{3}$, and A.M.~Rotunno$^{1,2}$}
\address{
		$^1$~Dipartimento di Fisica, Universit\`a di Bari, Via Amendola 173, 70126, Bari, Italy								\smallskip \\ 
		$^2$~Istituto Nazionale di Fisica Nucleare (INFN), Sezione di Bari,
		Via Orabona 4, 70126 Bari, Italy 										\smallskip \\
		$^3$~AHEP Group, Institut de F\'isica Corpuscular, CSIC/Universitat de Val\`encia,
		Edifici Instituts d'Investigaci\'o, Apt.\ 22085, 46071 Val\`encia, Spain 		
		\medskip\medskip}

\begin{abstract}
Nailing down the unknown neutrino mixing angle $\theta_{13}$ is one of the most 
important goals in current lepton physics. In this context,  
we perform a global analysis of neutrino oscillation data, focusing on $\theta_{13}$, 
and including recent
results  [{\em Neutrino 2008,
Proceedings of the XXIII International Conference on Neutrino Physics and Astrophysics,
Christchurch, New Zealand, 2008\/} (unpublished)]. We discuss two converging hints of
$\theta_{13}>0$, each at the level of $\sim\!\! 1\sigma$: an older one
coming from atmospheric neutrino data, and a newer one 
coming from the combination of solar 
and long-baseline reactor neutrino data. Their combination provides the global 
estimate $$\sin^2\theta_{13}=0.016\pm 0.010 \ (1\sigma)\,,$$ 
implying a preference for $\theta_{13}>0$ 
with non-negligible statistical significance ($\sim\!\!90\%$ C.L.). 
We discuss possible refinements of the experimental 
data analyses, which might sharpen such intriguing indication. 
\end{abstract}
 \maketitle


{\em Introduction.}---In the last decade, it has been established that
the neutrino states  $(\nu_e,\nu_\mu,\nu_\tau)$ with definite flavor are
quantum superpositions of states $(\nu_1,\nu_2,\nu_3)$ with
definite masses $(m_1,\,m_2,\,m_3)$ \cite{Kayser}. 
These findings point towards
new physics in the lepton sector, probably originating
at very high mass scales \cite{Moha}.

Independently of the origin of neutrino masses and mixing,
oscillation data can be accommodated in a simple
theoretical framework (adopted hereafter), where flavor and mass states are connected
by a
unitary mixing matrix $U$, parametrized in terms
of three mixing angles $(\theta_{12},\theta_{13},\theta_{23})$
and one CP-violating phase $\delta$ \cite{Kayser}. The mass spectrum gaps
can be parametrized in terms of $\delta m^2=m^2_2-m^2_1$ and
of $\Delta m^2=m^2_3-(m^2_1+m^2_2)/2$ \cite{Review}. 

Within this framework, the mass-mixing 
oscillation parameters
$(\delta m^2,\sin^2\theta_{12})$ and $(\Delta m^2,\sin^2\theta_{23})$
are rather well determined \cite{Review}. Conversely,  only upper bounds could be
placed so far on $\sin^2\theta_{13}$ ,
a dominant role being played by the null results of
the short-baseline CHOOZ reactor experiment \cite{CHOOZ} 
($\sin^2\theta_{13}\lesssim $~few\%). 

Determining a lower bound for 
$\theta_{13}$ (unless $\theta_{13}\equiv 0$ for some unknown reason)  
is widely recognized as a step of paramount importance in experimental
and theoretical neutrino
physics \cite{Kayser,Moha}.
Indeed, any future 
investigation of leptonic CP violation (i.e., of $\delta$), 
and of the neutrino mass spectrum hierarchy
[i.e., of sgn$(\Delta m^2)$] crucially depends on finding a nonzero value
for $\theta_{13}$. A worldwide program of direct
$\theta_{13}$ measurements with reactor and accelerator neutrinos is in 
progress, as recently reviewed, e.g., at the
recent {\em Neutrino 2008\/} Conference \cite{Nu2008}. 
 
In this context, any indirect indication in favor of $\theta_{13}>0$ becomes
highly valuable as a target for direct searches. We report here
two indirect, independent hints of $\theta_{13}>0$, one coming from
older atmospheric neutrino data, and one from the combination
of recent solar and long-baseline reactor data, as obtained by a 
global analysis of world oscillation searches. For the first time,
these hints add up to an overall indication in favor of $\theta_{13}>0$ at  
non-negligible confidence level of $\sim\!\!\,90\%$.

\medskip

{\em Hint from atmospheric neutrino data}.---In a 
previous analysis of world neutrino oscillation data \cite{Review}, we found
a weak hint in favor of $\theta_{13}>0$, at the level
of $\sim\!\! 0.9\sigma$, coming from atmospheric neutrino data
combined with accelerator and CHOOZ data (see Figs.~26 and
27 in \cite{Review}). We traced its origin  in
subleading $3\nu$ oscillation terms  driven by $\delta m^2$ \cite{Peres}, which are 
most effective at $\cos\delta=-1$ (see Fig.~24 in \cite{Review}), 
and which 
could partly explain the observed excess of sub-GeV atmospheric
electron-like events \cite{FN1}.
Such hint has persisted after combination with
further long-baseline (LBL) accelerator neutrino data \cite{Fogli2004,Fogli2006},
which have not yet placed strong constraints to $\nu_e$ appearance.
In particular, after including the Main Injector Neutrino Oscillation Search
(MINOS) data \cite{MINOS2008} presented at {\em Neutrino
2008\/} \cite{MINOS2008b}, and marginalizing over the leading mass-mixing
parameters $(\Delta m^2,\sin^2\theta_{23})$, we still find a $\sim \! 0.9\sigma$
hint of $\theta_{13}>0$ from the current combination of atmospheric, LBL accelerator, 
and CHOOZ data,
\begin{equation}
\sin^2\theta_{13}=0.012\pm 0.013 \ \ (1\sigma, \ \mathrm{Atm + LBL + CHOOZ}), 
\end{equation}
where the error scales almost linearly up to $\sim\!\! 3\sigma$, within the
physical range $\sin^2\theta_{13}\geq 0$.

\medskip

{\em Hint from solar and KamLAND data.}---In 
past years, the above ``atmospheric $\nu$ hint'' was not supported by 
independent long-baseline reactor
and solar neutrino data, which systematically preferred
$\theta_{13}=0$ as best fit, both separately and  in combination \cite{Review}. Therefore, in 
the global data analysis,
the  hint of $\theta_{13}>0$ was diluted 
well below $1\sigma$, and could be conservatively ignored \cite{Review}.

Such trend has recently changed, however, after the latest data release from the
Kamioka Liquid Scintillator Anti-Neutrino Detector (KamLAND) \cite{KL2008}, which favors
 slightly higher values of $\sin^2\theta_{12}$,
as compared to solar neutrino data 
\cite{SNO2} at fixed $\theta_{13}=0$.  
As discussed in \cite{NOVE}, and soon after in \cite{Baha}, this small difference
in $\sin^2\theta_{12}$ can be reduced for $\theta_{13}>0$, due to
the different dependence of the survival probability $P_{ee}=P(\nu_e\to\nu_e)$ on
the parameters ($\theta_{12},\theta_{13}$) 
for solar and KamLAND neutrinos \cite{Goswami}. Indeed,  
recent combinations of solar
and KamLAND data prefer $\theta_{13}>0$, although weakly 
\cite{NOVE,Baha,Valle}.

\begin{figure}
\includegraphics[width=0.9\textwidth]{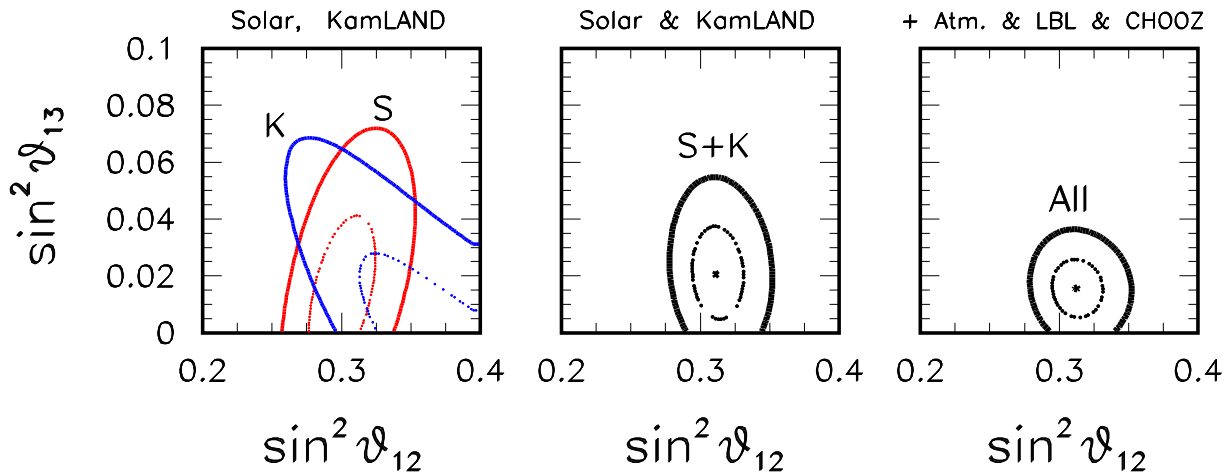}
\caption{\label{fig_1} 
Allowed regions in the plane $(\sin^2\theta_{12},\,\sin^2\theta_{13})$:
contours at $1\sigma$ (dotted) and $2\sigma$ (solid). Left and middle
panels: solar (S) and KamLAND (K) data, both separately (left)
and in combination (middle). In the left panel, the S contours are obtained 
by marginalizing the $\delta m^2$ parameter as constrained by KamLAND. 
Right panel: All data.}
\end{figure}

Remarkably, the recent
data from the third and latest phase of the Sudbury Neutrino Observatory (SNO)
\cite{SNO3} presented at {\em Neutrino 2008} \cite{SNO3b} further reduce 
the solar neutrino range for
$\sin^2 \theta_{12}$ and, in combination with KamLAND data, are thus expected to strengthen 
such independent hint in favor of $\theta_{13}>0$. We include
SNO-III data in the form of two new integral determinations of the 
charged-current (CC) and neutral current (NC) event rates \cite{SNO3}, with
error correlation $\rho\simeq -0.15$ inferred from the quoted CC/NC ratio
error \cite{SNO3}, but neglecting possible (so far unpublished) correlations
with previous SNO data \cite{SNO2}. We 
ignore the SNO-III elastic scattering (ES) event rate \cite{FN2}, which 
appears to be affected by statistical fluctuations \cite{SNO3,SNO3b} and which is, in any case, 
much less accurate than the solar neutrino ES rate measured by
Super-Kamiokande \cite{SKsolar}.

In the solar neutrino analysis, we update the total Gallium rate ($66.8\pm3.5$ SNU) \cite{FN3}
to account for a recent
reevaluation of the GALLEX data \cite{GALLEX,GALLEXb}.
The latest Borexino data \cite{Borexino,Ianni}, presented at {\em Neutrino 2008} 
\cite{Borexinob},
are also included for the sake of completeness. 
We do not include the 
Super-Kamiokande phase-II results \cite{SKII}, which would not provide 
significant additional constraints.
Finally, concerning KamLAND, we analyze
the full spectrum reported in \cite{KL2008}, and marginalize
away the low-energy geoneutrino fluxes from U and Th decay in the fit.
We have checked that our results agree well with
the published ones (in the case $\theta_{13}=0$) both 
on the oscillation parameters $(\delta m^2,\sin^2\theta_{12})$
and on the estimated geo-$\nu$ fluxes \cite{Rotunno}.

Figure~1 (left panel) shows the regions separately allowed
at $1\sigma$ ($\Delta\chi^2=1$, dotted) and $2\sigma$ ($\Delta\chi^2=4$, solid)
from the analysis of solar (S) and KamLAND (K)  neutrino data, in the
plane spanned by the mixing parameters $(\sin^2\theta_{12},\, \sin^2\theta_{13})$.
The $\delta m^2$ parameter is always marginalized away in  the KamLAND preferred region
(which is equivalent, in practice, to set $\delta m^2$ at its best-fit value
$7.67\times 10^{-5}$~eV$^2$).  The mixing parameters are 
positively and negatively correlated in the solar and KamLAND regions, respectively,
as a result of different functional forms for $P_{ee}(\sin^2\theta_{12},\sin^2\theta_{13})$
in the two cases.
The S and K allowed regions, which do not overlap at $1\sigma$
for $\sin^2 \theta_{13}=0$, merge for $\sin^2\theta_{13}\sim\mathrm{few}\times 10^{-2}$. 
The best fit (dot) and error ellipses (in black) for the solar+KamLAND  combination are shown
in the middle panel of Fig.~1. A hint of $\theta_{13}>0$ emerges at  $\sim\!\!1.2\sigma$ level,
\begin{equation}
\sin^2\theta_{13}=0.021\pm 0.017 \ \ (1\sigma, \ \mathrm{solar + KamLAND})\ ,
\end{equation}
with errors scaling linearly, to a good approximation, up to $\sim\!\!3\sigma$.  

\begin{figure}[t]
\includegraphics[width=0.97\textwidth]{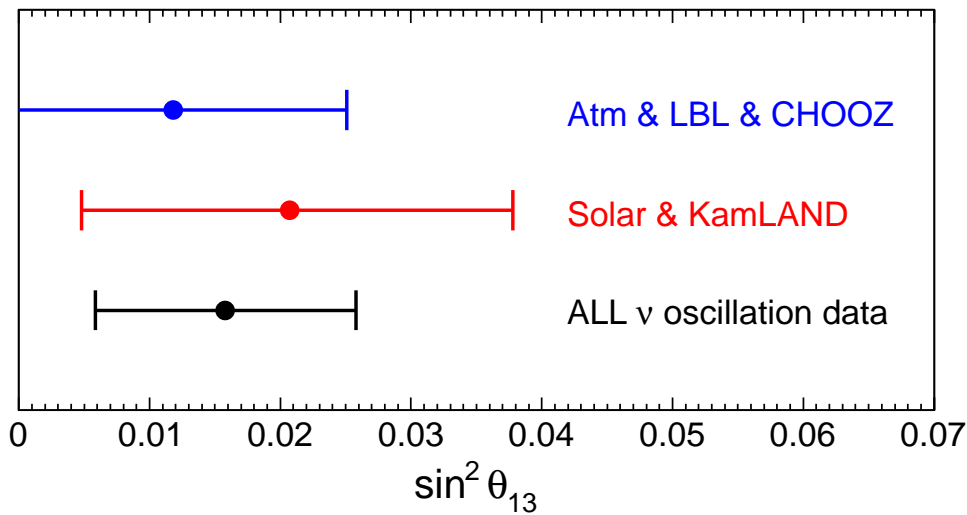}
\caption{\label{fig_2} 
Global $\nu$ oscillation analysis: Allowed $1\sigma$ ranges of 
$\sin^2\theta_{13}$ from different input data.} 
\end{figure}

\medskip
{\em Combination.} We have found two independent
hints of $\theta_{13}>0$, each at a level of
$\sim\!\! 1\sigma$, and with mutually consistent 
ranges for $\sin^2\theta_{13}$. Their combination
reinforces the overall preference for $\theta_{13}>0$, which emerges at the
level of $\sim\!\! 1.6\sigma$ in our global analysis. In particular, Fig.~1 (right panel)
shows the $1\sigma$ and $2\sigma$ error ellipses in the $(\sin^2\theta_{12},\, \sin^2\theta_{13})$ plane from the fit to all data,
which summarizes our current knowledge of electron neutrino mixing \cite{FN4}.
Marginalizing the $\sin^2\theta_{12}$ parameter we get
\begin{equation}
\sin^2\theta_{13}=0.016\pm 0.010 \ \ (1\sigma, \ \mathrm{all\ oscillation\ data})\ ,
\end{equation}
with linearly scaling errors. This is the most important result of our work.
Allowed ranges for other oscillation parameters are reported separately \cite{Addendum}. 
Summarizing, we find an overall preference for $\theta_{13}>0$ at
$\sim\!\!1.6\sigma$  or, equivalently, at $\sim\!\! 90\%$ C.L., from a global 
analysis of neutrino oscillation data, as available after the {\em Neutrino 2008\/} Conference.
The preferred $1\sigma$ ranges are summarized in Eqs.~(1)--(3), and are graphically
displayed in Fig.~2.

\medskip
{\em Conclusions and Prospects.}---In this Letter, we have focused on the last unknown neutrino
mixing angle $\theta_{13}$. Within a global analysis of world neutrino oscillation data, we have
discussed two hints in favor of $\theta_{13}>0$, each at the level of $\sim\!\! 1\sigma$.
Their combination provides an overall indication for $\theta_{13}>0$ at
a non-negligible 90\% confidence level.
To some extent, the present hints of $\theta_{13}>0$ can be corroborated
by more refined analyses. Concerning atmospheric neutrinos, an official, complete $3\nu$ analysis
by the Super-Kamiokande collaboration, including all experimental details, 
would be very important. The analysis should include $\delta m^2$-driven terms in
the oscillation probability \cite{Workshop,Ograms}, which have been neglected in the official 
publication \cite{SK3nu}. 
Concerning solar neutrinos, a detailed, fully documented and official
combination of all the SNO-I, II, and III
data \cite{Klein} would be helpful to sharpen the bounds on solar $\nu_e$ mixing and to contrast them
with (future) KamLAND data. The latter would benefit by a further reduction of the
normalization error, which is directly transferred to the mixing parameters.
In our opinion, such improvements might corroborate the statistical
significance of the previous hints by another $\sim\!\!1\sigma$ but, of course,
could not replace direct 
experimental searches for $\theta_{13}$ at reactors or accelerators. Two hints make for
a stronger indication, but do not make for a compelling proof.

\medskip\medskip
{\em Acknowledgments.}
G.L.F., E.L., A.M., and A.M.R.\ acknowledge 
support by the Italian MIUR and INFN through the ``Astroparticle Physics'' 
research project, and by the EU ILIAS through the ENTApP project. 
A.P.\ thanks J.W.F.~Valle for kind hospitality at IFIC, and acknowledges
support by MEC under the I3P program, by Spanish grants FPA2005-01269
and by European Commission network MRTN-CT-2004-503369 and
ILIAS/N6 RII3-CT-2004-506222.


\newpage
\begin{thebibliography}{99}

\bibitem{Kayser}
C.~Amsler {\em et al.}, Phys.\ Lett.\ B {\bf 667}, 1 (2008);
see also the review   ``Neutrino Mass, Mixing, and Flavor Change''
by B.~Kayser, {\em ibid.} {\bf 667}, 163 (2008).


\bibitem{Moha}
 R.~N.~Mohapatra and A.~Y.~Smirnov,
 ``Neutrino mass and new physics,''
  Ann.\ Rev.\ Nucl.\ Part.\ Sci.\  {\bf 56}, 569 (2006).
   
\bibitem{Review}
 G.~L.~Fogli, E.~Lisi, A.~Marrone and A.~Palazzo,
  ``Global analysis of three-flavor neutrino masses and mixings,''
  Prog.\ Part.\ Nucl.\ Phys.\  {\bf 57}, 742 (2006)
  [arXiv:hep-ph/0506083].

\bibitem{CHOOZ}
  M.~Apollonio {\it et al.}  [CHOOZ Collaboration],
  ``Search for neutrino oscillations on a long base-line at the CHOOZ  nuclear
  power station,''
  Eur.\ Phys.\ J.\  C {\bf 27}, 331 (2003)
  [arXiv:hep-ex/0301017].
  
\bibitem{Nu2008} 
         {\em Neutrino 2008}, XXIII International Conference on 
		Neutrino Physics and Astrophysics (Christchurch, New Zealand, 2008).
		Webpage: {\tt www2.phys.canterbury.ac.nz/$\sim$jaa53} 

\bibitem{Peres}
  O.~L.~G.~Peres and A.~Y.~Smirnov,
  ``Testing the solar neutrino conversion with atmospheric neutrinos,''
  Phys.\ Lett.\  B {\bf 456}, 204 (1999)
  [arXiv:hep-ph/9902312];    O.~L.~G.~Peres and A.~Y.~Smirnov,
  ``Atmospheric neutrinos: LMA oscillations, $U_{e3}$ induced interference and CP-violation,''
  Nucl.\ Phys.\  B {\bf 680}, 479 (2004)
  [arXiv:hep-ph/0309312].

\bibitem{FN1}
Another refined $3\nu$
analysis of atmospheric $\nu$ data has not found an appreciable 
preference for $\theta_{13}>0$:
M.~C.~Gonzalez-Garcia and M.~Maltoni,
  ``Phenomenology with Massive Neutrinos,''
 Phys.\ Rept.\  {\bf 460}, 1 (2008)
  [arXiv:0704.1800 [hep-ph]]. On the other hand, a recent (although less documented) 
analysis seems indeed to favor $-\cos\delta\sin \theta_{13}>0$ as in our case:
J.~Escamilla, D.~C.~Latimer and D.~J.~Ernst,
  ``Atmospheric neutrino oscillation data constraints on $\theta_{13}$,''
  arXiv:0805.2924 [nucl-th].


\bibitem{Fogli2004}
  G.~L.~Fogli, E.~Lisi, A.~Marrone, A.~Melchiorri, A.~Palazzo, P.~Serra and J.~Silk,
  ``Observables sensitive to absolute neutrino masses: Constraints and
  correlations from world neutrino data,''
  Phys.\ Rev.\  D {\bf 70}, 113003 (2004)
  [arXiv:hep-ph/0408045].

\bibitem{Fogli2006}
  G.~L.~Fogli, E.~Lisi, A.~Marrone, A.~Melchiorri, A.~Palazzo, P.~Serra, J.~Silk, and A.~Slosar,
  ``Observables sensitive to absolute neutrino masses: A reappraisal after
  WMAP-3y and first MINOS results,''
  Phys.\ Rev.\  D {\bf 75}, 053001 (2007)
  [arXiv:hep-ph/0608060].

\bibitem{MINOS2008}
  MINOS Collaboration, P.~Adamson {\em et al.},
  ``Measurement of Neutrino Oscillations with the MINOS Detectors in the NuMI Beam,''
  arXiv:0806.2237 [hep-ex]; to be published in Phys.\ Rev.\ Lett.
  
\bibitem{MINOS2008b} 
   H.~Gallagher, in {\em Neutrino 2008\/} \protect\cite{Nu2008}.

\bibitem{KL2008}
  KamLAND Collaboration, S.~Abe {\it et al.},
  ``Precision Measurement of Neutrino Oscillation Parameters with KamLAND,''
  Phys.\ Rev.\ Lett.\ {\bf 100}, 221803
  [arXiv:0801.4589 [hep-ex]].

\bibitem{SNO2}
	SNO Collaboration,
  B.~Aharmim {\it et al.},
  ``Electron energy spectra, fluxes, and day-night asymmetries of $^8$B solar
  neutrinos from the 391-day salt phase SNO data set,''
  Phys.\ Rev.\  C {\bf 72}, 055502 (2005)
  [arXiv:nucl-ex/0502021].

\bibitem{NOVE} G.L.\ Fogli, E.\ Lisi, A.\ Marrone, A.\ Palazzo, and A.M.~Rotunno, in Proceedings
  of {\em NO-VE 2008}, IV International Workshop on
``Neutrino Oscillations in Venice'' (Venice, Italy, April 15-18, 2008),
 edited by M.~Baldo Ceolin (University of Padova, Papergraf Editions, Padova, Italy, 2008), p.~21;
 also available at:
{\tt neutrino.pd.infn.it/NO-VE2008}

\bibitem{Baha}
  A.~B.~Balantekin and D.~Yilmaz,
  ``Contrasting solar and reactor neutrinos with a non-zero value of $\theta_{13}$,''
  J.\ Phys.\ G {\bf 35}, 075007 (2008)
  [arXiv:0804.3345 [hep-ph]].

\bibitem{Goswami}
  S.~Goswami and A.~Y.~Smirnov,
  ``Solar neutrinos and 1-3 leptonic mixing,''
  Phys.\ Rev.\  D {\bf 72}, 053011 (2005)
  [arXiv:hep-ph/0411359].

\bibitem{Valle}
  M.~Maltoni, T.~Schwetz, M.~A.~Tortola and J.~W.~F.~Valle,
  ``Status of global fits to neutrino oscillations,''
  New J.\ Phys.\  {\bf 6}, 122 (2004)
  [arXiv:hep-ph/0405172 v6].
  
\bibitem{SNO3}
  SNO Collaboration, 
  B.~Aharmim {\it et al.},
  ``An Independent Measurement of the Total Active $^8$B Solar Neutrino Flux Using
  an Array of $^3$He Proportional Counters at the Sudbury Neutrino Observatory,''
  Phys.\ Rev.\ Lett.\ {\bf 101}, 111301 (2008)
  [arXiv:0806.0989 [nucl-ex]].


\bibitem{SNO3b}
	H.~Robertson, in {\em Neutrino 2008\/} \protect\cite{Nu2008}.


\bibitem{FN2} We reproduce well the official SNO-III results \protect\cite{SNO3}
for the mass-mixing parameter ranges and best-fit values (at $\theta_{13}=0$). 
We have also checked that these ranges are not altered by including the
SNO-III ES rate, which merely increases the overall $\chi^2$ by $\sim\!\! 3.5$ in our fit. 

    
\bibitem{SKsolar}
 Super-Kamiokande Collaboration,
 J.~Hosaka {\it et al.},
  ``Solar neutrino measurements in Super-Kamiokande-I,''
  Phys.\ Rev.\  D {\bf 73}, 112001 (2006)
  [arXiv:hep-ex/0508053].
  

\bibitem{FN3}
The  comparison of Gallium and SNO data is crucial to
bound  $\sin^2\theta_{13}$ with  solar $\nu$ data
only \protect\cite{Review,Goswami}.

\bibitem{GALLEX} F.~Kaether, 
	``Data Analysis of the solar neutrino experiment GALLEX''
	(PhD Thesis, Heidelberg, 2007).


\bibitem{GALLEXb} R.L.~Hahn, in {\em Neutrino 2008\/} \protect\cite{Nu2008}.

  
\bibitem{Borexino}   Borexino Collaboration, C.~Arpesella {\em et al.,} 
  ``New results on solar neutrino fluxes from 192 days of Borexino data,''
  arXiv:0805.3843 [astro-ph].

\bibitem{Ianni} We thank A.~Ianni for useful information about the latest Borexino data.

\bibitem{Borexinob} C.~Galbiati, in {\em Neutrino 2008\/} \protect\cite{Nu2008}.

\bibitem{SKII} Super-Kamiokande Collaboration,
  J.~P.~Cravens {\it et al.},
  ``Solar neutrino measurements in Super-Kamiokande-II,''
  Phys.\ Rev.\ D {\bf 78}, 032002 (2008)
  [arXiv:0803.4312 [hep-ex]].

\bibitem{Rotunno}
	G.L.~Fogli, E.~Lisi, A.~Palazzo, and A.~M.~Rotunno, preprint in preparation.
	Preliminary geo-$\nu$ results from this work have been kindly
	shown by J.G.~Learned in his talk  at {\em Neutrino 2008\/} \protect\cite{Nu2008}.


\bibitem{FN4} The $\nu_e$ mixing with $\nu_i$ is determined by $|U_{e1}|^2=\cos^2\theta_{13}\cos^2\theta_{12}$,
$|U_{e2}|^2=\cos^2\theta_{13}\sin^2\theta_{12}$, and $|U_{e3}|^2=\sin^2\theta_{13}$.


\bibitem{Addendum} G.~L.~Fogli {\it et al.},
  ``Observables sensitive to absolute neutrino masses.~II,''
  Phys.\ Rev.\ D {\bf 78}, 033010 (2008)
    [arXiv:0805.2517 [hep-ph]].


\bibitem{Workshop}
	See the contributions at the {\em International Workshop on sub-dominant oscillation effects in atmospheric neutrino experiments}, Research Center for Cosmic Neutrino (Kashiwa, Japan, 2004), edited by
T.~Kajita and K.~Okumura, Frontier Science Series n. {\bf 45}, 
(Universal Academy Press, Tokyo, Japan, 2005). Webpage:
{\tt www-rccn.icrr.u-tokyo.ac.jp/rccnws04}

\bibitem{Ograms}
  E.~K.~Akhmedov, M.~Maltoni and A.~Y.~Smirnov,
  ``Neutrino oscillograms of the Earth: effects of 1-2 mixing and
  CP-violation,''
  J.\ High Energy Phys.\ 06 (2008) 072
  [arXiv:0804.1466 [hep-ph]].

\bibitem{SK3nu}
  Super-Kamiokande Collaboration, J.~Hosaka {\it et al.},
  ``Three flavor neutrino oscillation analysis of atmospheric neutrinos in
  Super-Kamiokande,''
  Phys.\ Rev.\  D {\bf 74}, 032002 (2006)
  [arXiv:hep-ex/0604011].


\bibitem{Klein}
   The SNO Collaboration members are planning a combined analysis of all their data with 
   the lowest 
   possible analysis threshold, see J.~Klein,
   Colloquium on ``Results and Prospects with the Sudbury Neutrino Observatory''
   (NIKHEF, Amsterdam, The Netherlands, 2008); available at:
   {\tt http://agenda.nikhef.nl/conferenceDisplay.py?confId=249}. See also the
   talk by A.~McDonald, ``SNO and the New SNOLAB Underground Facility'' at
   {\em PASCOS~'08}, 4th International Symposium on Particles, Strings and Cosmology
   (Perimeter Institute, Waterloo, Ontario, Canada, 2008); available at
   {\tt http://pirsa.org/08060047 }
   


\end{thebibliography}
\end{document}